\documentclass[aps,pra,reprint,superscriptaddress,floatfix]{revtex4-2}

\usepackage[colorlinks=true,urlcolor=blue,citecolor=magenta]{hyperref}
\usepackage{amsmath,amssymb,bm,bbm}
\usepackage{graphicx}
\usepackage{float}
\usepackage{cancel}
\usepackage{physics}
\usepackage{braket}
\usepackage{enumitem}
\usepackage[normalem]{ulem}

\usepackage{tcolorbox}

\begin{document}

\title{Quasiparticle projection method for dynamically unstable Bose--Einstein condensates}

\author{Asier Izquierdo}
\affiliation{Department of Physical Chemistry, University of the Basque Country UPV/EHU, 48080 Bilbao, Spain}
\affiliation{EHU Quantum Center, University of the Basque Country UPV/EHU, Leioa, Biscay, Spain}

\author{Maria Arazo}
\affiliation{Department of Physics, University of the Basque Country UPV/EHU, 48080 Bilbao, Spain}
\affiliation{EHU Quantum Center, University of the Basque Country UPV/EHU, Leioa, Biscay, Spain}

\author{Sofía Martínez-Garaot}
\affiliation{Department of Physical Chemistry, University of the Basque Country UPV/EHU, 48080 Bilbao, Spain}
\affiliation{EHU Quantum Center, University of the Basque Country UPV/EHU, Leioa, Biscay, Spain}

\author{Michele Modugno}
\affiliation{Department of Physics, University of the Basque Country UPV/EHU, 48080 Bilbao, Spain}
\affiliation{IKERBASQUE, Basque Foundation for Science, 48013 Bilbao, Spain}
\affiliation{EHU Quantum Center, University of the Basque Country UPV/EHU, Leioa, Biscay, Spain}

\begin{abstract}
We present a general formalism for performing a time-dependent Bogoliubov analysis of a dynamically unstable Bose--Einstein condensate, which extends the quasiparticle projection method of Morgan \textit{et al.} [\href{https://doi.org/10.1103/PhysRevA.57.3818}{Phys.\ Rev.\ A \textbf{57}, 3818 (1998)}] to cases with a complex spectrum. By introducing the proper left eigenvectors associated with each regime, we construct a biorthogonal basis. While the usual Bogoliubov normalization $\braket{u | u} - \braket{v | v} = 1$ may not hold in this basis, it still allows for a complete mode decomposition and an accurate reconstruction of arbitrary perturbations over time.
This approach extends the applicability of the Bogoliubov framework beyond the stable regime, providing a consistent analysis of the time evolution of unstable condensates.
As a proof of concept, we apply the method to a one-dimensional condensate with attractive interactions, which is dynamically unstable and evolves into nonstationary localized structures seeded by small perturbations.
Overall, the present method provides a complete and robust mode expansion that remains meaningful beyond the linear regime and useful for characterizing the macroscopic development of instabilities.
\end{abstract}

\maketitle

\section{Introduction}

Bogoliubov theory provides a general framework for analyzing the stability and elementary excitations of many-body systems governed by nonlinear equations, such as the Gross--Pitaevskii (GP) equation~\cite{fetter1972,dalfovo1999,stringari_book}.
To analyze a system's response to fluctuations, the dynamics are linearized around a known stationary solution by introducing small perturbations and studying their time evolution. 
This leads to the Bogoliubov–de Gennes equations~\cite{castin2001}, which determine the system’s excitation spectrum---a key ingredient for understanding the collective dynamics of Bose--Einstein condensates (BECs) and their instability mechanisms under external perturbations. 
If all excitation frequencies are real, the system is dynamically stable, and perturbations oscillate without growing. In contrast, the presence of imaginary components in the spectrum signals a dynamical instability, characterized by perturbations that grow exponentially over time.

In the stable regime, Bogoliubov theory yields a well-defined mode expansion. Although the Bogoliubov--de Gennes equations represent a non-Hermitian eigenvalue problem, a complete and orthonormal set of modes can still be constructed when the spectrum is purely real, ensuring a well-defined time evolution~\cite{morgan1998,castin2001}. 
However, when the spectrum acquires imaginary parts (indicating instability), the standard mode expansion breaks down. 
In this regime, the Bogoliubov norm $\braket{u|u} - \braket{v|v}$ becomes zero, making the time evolution ill-defined within the conventional Bogoliubov framework~\cite{wu2001,wu2003}.
For these reasons, it is widely believed that Bogoliubov theory loses its validity in the unstable regime, restricting its applicability primarily to predicting the onset of instability---that is, the growth rates of the unstable modes around the initial stationary solution.

To overcome this apparent difficulty, Leonhardt \textit{et al.}~\cite{leonhardt2003} proposed to maintain the standard normalization while abandoning the single-frequency Bogoliubov modes by defining suitable combinations of different modes. This approach has been employed, for example, to discuss quantum hydrodynamic instabilities, such as those emerging in the dynamics of sonic horizons within BECs~\cite{leonhardt2003,jain2007} (see also Refs.~\cite{garay2000,garay2001,ribeiro2022}). 
A similar strategy, which abandons the single-quantum-number basis while retaining the standard Bogoliubov normalization, was introduced by Kawaguchi and Ohmi \cite{kawaguchi2004} in the context of splitting instabilities of multiply quantized vortices. 
From a complementary quantum-field theoretical perspective, Mine \textit{et al.}~\cite{mine2007} demonstrated that the presence of complex Bogoliubov--de Gennes eigenvalues implies an indefinite-metric state space, requiring the introduction of physical-state conditions to describe metastable condensates and recover unstable dynamics within linear response theory. This body of work establishes a consistent quantization framework for dynamically unstable condensates, but primarily focuses on diagnosing stability or describing the onset of instabilities, rather than reconstructing the full nonlinear time evolution of the condensate wave function.

Here we follow a different approach, faithful to the original Bogoliubov expansion, in the sense that we construct a biorthogonal basis of single-frequency modes. This enables a complete mode decomposition with the usual normalization condition, $\braket{L_j | R_{j’}} = \delta_{j,j’}$, by employing the appropriate definitions of right~$|R_j\rangle$ and left~$|L_j\rangle$ eigenvectors. This way, we obtain a clearer formal extension of Bogoliubov theory that remains valid even in the presence of dynamical instabilities, where the excitation spectrum includes complex or purely imaginary eigenvalues.
We use this framework to generalize the \emph{quasiparticle projection method} introduced by Morgan \textit{et al.}~\cite{morgan1998}, enabling the analysis of systems undergoing instabilities even beyond the linear regime of small perturbations.
This approach extends the applicability of the Bogoliubov framework away from the stable regime, providing a consistent analysis of the time evolution of unstable condensates and enabling an accurate reconstruction of the condensate wave function over time under arbitrary perturbations.
It has potential applications to a wide range of scenarios in ultracold gases and beyond, particularly when a stabilizing mechanism prevents collapse and drives the instability toward emergent behaviors, such as the formation of droplets or supersolids. 
Here, as a proof of concept, we apply the method to a one-dimensional attractive BEC, where the balance between quantum pressure and attractive nonlinearity stabilizes the system~\cite{dalfovo1999}, supporting solitonic solutions \cite{pethick2008} and preventing collapse from dynamical instabilities. 

The method developed here offers a key conceptual insight.
While the usual Bogoliubov stability analysis is used to determine the onset of an instability and assumes the linear regime, where unstable modes are initially underpopulated relative to the condensate, the instability only leads to a modulation of the condensate once these modes become macroscopically populated and the system enters the nonlinear regime.
To overcome this conceptual gap, the present approach builds on the linear stability analysis, providing a methodological framework that enables tracking both the growth of the mode populations and their coupled dynamics, fully characterizing how the instability develops at the macroscopic level.

The paper is organized as follows.  
In Sec.~\ref{sec:GP&B}, we establish the notation and set up the model framework by briefly reviewing the GP equation and the Bogoliubov formalism for a BEC.  
We then review the spectral properties of the Bogoliubov matrix (Sec.~\ref{sec:spectral_properties}) and discuss the expansion in the basis of Bogoliubov modes (Sec.~\ref{sec:eigenbasis_expansion}).  
In Sec.~\ref{sec:realL}, we address the specific cases of real (Sec.~\ref{sec:real_spectrum}) and imaginary (Sec.~\ref{sec:imaginary_spectrum}) spectra.  
In Sec.~\ref{sec:qpm}, we generalize the quasiparticle projection method~\cite{morgan1998} to the case of a generic complex spectrum, discussing first the general framework (Sec.~\ref{sec:general_case}) and then the particular case in which the spectrum has the form $\varepsilon = \pm \sqrt{f}$, with $f$ a real-valued function that can take either positive or negative values (Sec.~\ref{sec:real_imaginary_spectrum}).  
As a proof of concept, we apply the method to an attractive BEC in a box in Sec.~\ref{sec:proof_of_concept}, considering both zero (Sec.~\ref{sec:bec_at_rest}) and nonzero (Sec.~\ref{sec:moving_bec}) velocities.  
Finally, we summarize our results and draw conclusions in Sec.~\ref{sec:conclusions}.
The two Appendices briefly discuss alternative factorizations of stationary and complete solutions (Appendix~\ref{app:alt_factorization}), and the orthogonalization of fluctuations to the zero-energy mode (Appendix~\ref{app:zeroenergy}).

\section{Gross--Pitaevskii equation and Bogoliubov formalism}
\label{sec:GP&B}

To establish the notation and set up the model framework, we consider a BEC described by the wave function $\phi(\bm{r}, t)$, which evolves according to the GP equation,
\begin{equation} \label{eq:GPE}
	i \hbar \partial_t \phi(\bm{r}, t)\!=\!\left[-\frac{\hbar^{2}}{2 m} \nabla^{2}+V(\bm{r}) + g|\phi(\bm{r}, t)|^{2}\right]\! \phi(\bm{r}, t) ,
\end{equation}
where the density is normalized to the total number of particles in the condensate,
$ \int |\phi(\bm{r},t)|^2 d\bm{r} = N$. To introduce the standard Bogoliubov theory~\cite{fetter1972,castin2001}, we consider a stationary solution of the above equation, $\phi(\bm{r},t) = \phi_{0}(\bm{r}) \exp{-i\mu t/\hbar}$, with chemical potential $\mu$, and introduce small deviations  $\delta\phi(\bm{r},t)$, writing
\begin{equation} \label{eq:small_deviation_from_steady}
\phi(\bm{r},t)	= e^{-i\mu t/\hbar} \left[\phi_{0}(\bm{r}) + \delta{\phi}(\bm{r},t) \right].
\end{equation}
By inserting the above expression into the GP equation \eqref{eq:GPE} and neglecting the second-order terms in $\delta\phi$, one obtains 
\begin{equation}\label{eq:linear_system_delta_phi_L_GP}
	i\hbar\partial_t
	\begin{pmatrix}
		\delta\phi\\
		\delta\phi^{*}
	\end{pmatrix}=\mathcal{L}
	\begin{pmatrix}
		\delta\phi\\
		\delta\phi^{*}
	\end{pmatrix},
\end{equation}
where $\mathcal{L}$ is the following linear operator,
\begin{equation} \label{eq:L_GP}
	\mathcal{L}=
	\begin{pmatrix}
		H_{\textrm{GP}}+g|\phi_0|^2 & g\phi_0^2\\
		-g\phi_0^{*2} & -\left[H_{\textrm{GP}}+g|\phi_0|^2\right]^*
	\end{pmatrix}
	 \,,
\end{equation}
and $H_{\textrm{GP}}$ is the Hamiltonian
\begin{equation} \label{eq:H_GP}
	H_{\textrm{GP}}\equiv -\frac{\hbar^2}{2m}\nabla^{2} + V(\bm{r}) + g|\phi_0|^2-\mu\,.
\end{equation}
The potential $V$ is assumed to be real, ensuring that $H_{\textrm{GP}}$ is a Hermitian operator, i.e., $H_{\textrm{GP}}^\dagger = H_{\textrm{GP}}$. 
In the following, we also assume $\mathcal{L}$ is diagonalizable---a situation that occurs in many relevant configurations within Gross--Pitaevskii theory (see, e.g., Refs.~\cite{fetter1972,morgan1998,castin2001}).

We note that other factorizations of the condensate wave function, alternative to Eq.~\eqref{eq:small_deviation_from_steady}, are also possible and used in the literature. They are briefly reviewed in Appendix~\ref{app:alt_factorization}, along with the corresponding forms of the operator $\mathcal{L}$.

\subsection{Spectral properties of $\mathcal{L}$} 
\label{sec:spectral_properties}

Now, let us consider the eigenvalue equation for $\mathcal{L}$, with eigenvector $\left(u_{j},\, v_{j}\right)^\top$ and eigenvalue $\varepsilon_{j}$,
\begin{equation} \label{eq:eigenvector_eigenvalue_epsilon} 
	\mathcal{L}
	\begin{pmatrix}
		u_{j}\\
		v_{j}
	\end{pmatrix}=
	\varepsilon_{j}
	\begin{pmatrix}
		u_{j}\\
		v_{j}
	\end{pmatrix}
	 \,,
\end{equation}
whose solutions determine the so-called Bogoliubov functions $u_{j}(\bm{r})$ and $v_{j}(\bm{r})$.
In general, it can be shown that $\varepsilon_{j}^{*}$ is also an eigenvalue of $\mathcal{L}$ \cite{castin2001}.
In addition to this, the following spectral properties hold:
\begin{enumerate}
	
	\item \label{item:p2} $\left(v_{j}^{*},\, u_{j}^{*}\right)^\top$ is an eigenvector of $\mathcal{L}$, with eigenvalue $-\varepsilon_{j}^{*}$.
	
	\item \label{item:p3} If $H_{\textrm{GP}}\in\mathbb{R}$, then $\left(u_{j},\, -v_{j}\right)^\top$ is an eigenvector of $\mathcal{L}^\dagger$, with eigenvalue $\varepsilon_{j}$. 

    \item \label{item:p1} If $\mathcal{L}\in\mathbb{R}$, then $\left(u_{j}^{*},\, v_{j}^{*}\right)^\top$ is also an eigenvector of $\mathcal{L}$, with eigenvalue $\varepsilon_{j}^{*}$.

	\item \label{item:p4} If $\mathcal{L}\in\mathbb{R}$, then $\left(-v_{j}^{*},\, u_{j}^{*}\right)^\top$ is an eigenvector of $\mathcal{L}^\dagger$, with eigenvalue $-\varepsilon_{j}^{*}$.
\end{enumerate}

\subsection{Expansion in the eigenbasis of $\mathcal{L}$}
\label{sec:eigenbasis_expansion}

In general, it is useful to express the deviation $\delta \phi$ from the stationary state in terms of the eigenvectors of the Bogoliubov linear operator $\mathcal{L}$, as this allows us to separate the system’s fluctuations into normal modes that evolve independently. This approach simplifies the analysis of the dynamics, particularly when studying excitations and their time evolution.

We recall that \emph{normal matrices} are matrices that commute with their Hermitian conjugate, $[A, A^\dagger] = 0$, and have a set of complete (i.e., sufficient to span the entire vector space), orthogonal eigenvectors. 
In contrast, the eigenvectors of non-normal matrices---if non-degenerate and thus diagonalizable---satisfy a different orthogonality relation: biorthogonality between the left and right eigenvectors~\footnote{See, e.g., Chapter 11 of Ref.~\cite{press2007}.}.
Specifically, the left and right eigenvectors of a diagonalizable, non-normal operator $\mathcal{M}$ satisfy the bicompleteness relation \cite{castin2001},
\begin{equation}\label{eq:completeness}
    \sum_{j} \ket{\mathrm{R}_{j}}\bra{\mathrm{L}_{j}} = \mathbbm{1} \,,
\end{equation}
where $\ket{\mathrm{R}_{j}}$ is a right eigenvector of $\mathcal{M}$ with eigenvalue $m_{j}$,
\begin{equation} \label{eq:right_eigenvector}
    \mathcal{M}\ket{\mathrm{R}_{j}} = m_{j}\ket{\mathrm{R}_{j}} \,,
\end{equation}
and $\bra{\mathrm{L}_{j}}$ is a left eigenvector of $\mathcal{M}$ with the same eigenvalue $m_{j}$,
\begin{equation} \label{eq:left_eigenvector}
   \bra{\mathrm{L}_{j}} \mathcal{M} = m_{j}\bra{\mathrm{L}_{j}} \,.
\end{equation}
Equivalently, applying the Hermitian adjoint to both sides of Eq.~\eqref{eq:left_eigenvector}, one gets
\begin{equation} \label{eq:M_left_eigenvector}
    \mathcal{M}^{\dagger}\ket{\mathrm{L}_{j}} = m_{j}^{*}\ket{\mathrm{L}_{j}} \,.
\end{equation}
The sum over $j$ in Eq.~\eqref{eq:completeness} runs over all eigenvalues $m_{j}$ [see Eqs.~\eqref{eq:right_eigenvector} and~\eqref{eq:left_eigenvector}]. 
Finally, left and right eigenvectors are normalized so that they satisfy the biorthogonality condition
\begin{equation} \label{eq:biorthogonality}
    \braket{\mathrm{L}_{j} | \mathrm{R}_{j^\prime}} = \delta_{j, j^{\prime}} \,.
\end{equation}
Owing to the above properties, any state $\ket{\Psi}$ can be represented as~\cite{mayer2003} 
\begin{equation} \label{eq:expansion_left_right_eigenvec}
	\ket{\Psi} = \sum_{j} \left|{\mathrm{R}_{j}}\right\rangle \braket{\mathrm{L}_{j} | \Psi} = \sum_{j} b_{j} \ket{\mathrm{R}_{j}}, 
\end{equation}
with $b_{j} \equiv \braket{\mathrm{L}_{j} | \Psi}$.

Customarily, right eigenvectors of the Bogoliubov operator $\mathcal{L}$ are expressed in terms of the components $u$ and $v$, as in Eq.~\eqref{eq:eigenvector_eigenvalue_epsilon},
\begin{equation}
\label{eq:psiR}
\ket{\mathrm{R}_{j}} = \begin{pmatrix} \ket{u_{j}} \\ \ket{v_{j}} \end{pmatrix}.
\end{equation}
Similarly, left eigenvectors can be written as
\begin{equation}
\label{eq:psiL}
\ket{\mathrm{L}_{j}} = \begin{pmatrix} \ket{u_{j}^\textrm{L}} \\[2pt]  \ket{v_{j}^\textrm{L}} \end{pmatrix}.
\end{equation}
Then, the above biorthogonality condition \eqref{eq:biorthogonality} can be cast in the form
\begin{equation} \label{eq:biorthogonality2}
    \braket{{u}_{j}^L | {u}_{j'}} + \braket{{v}_{j}^L | {v}_{j'}} = \delta_{j,j'},
\end{equation}
and, for a vector such as $\ket{\delta\Phi}=\left(\ket{\delta\phi},\, \ket{\delta\phi^{*}}\right)^\top$, the coefficients $b_{j}$ in Eq.~\eqref{eq:expansion_left_right_eigenvec} take the form
\begin{equation}
    b_{j}  = \braket{u_{j}^\textrm{L} | \delta\phi} + \braket{v_{j}^\textrm{L} | \delta\phi^{*}}.
    \label{eq:bj}
\end{equation}
In general, unlike some specific cases---such as when $\mathcal{L} \in \mathbb{R}$, as we consider in the following section---it is not possible to obtain an explicit correspondence between left and right eigenvectors; namely, one cannot express $(\ket{u^\textrm{L}},\,\ket{v^\textrm{L}})$ in terms of $(\ket{u},\,\ket{v})$.

We also recall that, since vectors of the form $\ket{\delta\Phi}=\left(\ket{\delta\phi},\, \ket{\delta\phi^{*}}\right)^\top$ are invariant under the \emph{conjugation} transformation $\ket{\bar{\delta\Phi}}\equiv \sigma_{x} \ket{\delta\Phi^{*}}$ \cite{leonhardt2003}, the above expansion \eqref{eq:expansion_left_right_eigenvec} is conventionally written in the manifestly invariant form 
\begin{equation}
    \ket{\delta\Phi} = \sum_{j} b_{j} \begin{pmatrix} \ket{u_{j}} \\ \ket{v_{j}} \end{pmatrix} 
    + b_{j}^{*} \begin{pmatrix} \ket{v_{j}^{*}} \\ \ket{u_{j}^{*}} \end{pmatrix}.
    \label{eq:conventional_expansion}
\end{equation}
Note that, according to the spectral property~\ref{item:p2}, the second term corresponds to the eigenvectors of $\mathcal{L}$ with eigenvalues $-\varepsilon_{j}^{*}$. Since, in principle, the index $j$ runs over all possible eigenvalues $\varepsilon_j$, this must be taken into account either by renormalizing the coefficients $b_j$ in Eq.~\eqref{eq:bj} by a factor of $1/2$, or by restricting the sum to avoid double-counting the eigenvalues. The latter solution is often implicitly adopted in the literature.

In the following, we will see that alternative decompositions---which break manifest invariance but remain invariant---can be more convenient for dealing with situations in which the spectrum is not real. In any case, we remark that the expansions \eqref{eq:expansion_left_right_eigenvec} and \eqref{eq:conventional_expansion} are both general and complete.

\section{Special case: $\mathcal{L}\in\mathbb{R}$}
\label{sec:realL}

In this section, we focus on the case of a real Bogoliubov operator $\mathcal{L}$ for which, in particular, the spectral properties~\ref{item:p1} and~\ref{item:p4} also apply.
    
\subsection{Real spectrum}
\label{sec:real_spectrum}

When the spectrum is real, the Bogoliubov theory for BECs is well established and widely employed in the literature \cite{fetter1972,castin2001,dalfovo1999}. 
As discussed earlier, based on the spectral properties, the right eigenvectors can be organized into two families,
\begin{equation}
\ket{\mathrm{R}_{k}} = \begin{pmatrix} \ket{u_{k}} \\ \ket{v_{k}} \end{pmatrix},
\, \begin{pmatrix} \ket{v_{k}^{*}} \\ \ket{u_{k}^{*}} \end{pmatrix} \,,
\end{equation}
which, in this case, correspond to eigenvalues $\pm\varepsilon_{k}$, respectively (see spectral property~\ref{item:p2}). 
Here we set $+\varepsilon_{k}$ to represent the positive eigenvalues, arranged in ascending order. The ordering of $-\varepsilon_{k}$ follows accordingly. Note that, with this choice, the index $k$ runs over a subset of the eigenvalues (the positive ones) unlike the previous index $j$, which runs over all eigenvalues.

The corresponding left eigenvectors are (see spectral properties~\ref{item:p3} and \ref{item:p4})
\begin{equation}
\ket{\mathrm{L}_{k}} = \begin{pmatrix} \ket{u_{k}} \\ -\ket{v_{k}} \end{pmatrix},
\, \begin{pmatrix} -\ket{v_{k}^{*}} \\ \ket{u_{k}^{*}} \end{pmatrix}.
\end{equation}
They obey the well-known normalization condition~\cite{fetter1972,morgan1998,castin2001}
\begin{equation} \label{eq:normalization_condition_real_eigenvalue}
	\braket{\mathrm{L}_{k} | \mathrm{R}_{k^\prime}}
	= \braket{ u_{k} | u_{k'}} - \braket{ v_{k} | v_{k'}} = \delta_{k,k'}\,
\end{equation}
for both $\pm\varepsilon_{k}$. 
Then, the expansion for $\delta\phi$ can be written as
\begin{equation} \label{eq:delta_phi_real_espectra_first_row}
		\delta\phi(\bm{r},t)=\sum_{k}b_{k}(t)u_{k}(\bm{r})+b^*_{k}(t)v_{k}^{*}(\bm{r}) \,,
\end{equation}
where 
\begin{equation}
b_{k}(t) \equiv \braket{u_{k} | \delta\phi} - \braket{v_{k} | \delta\phi^{*}}. 
\end{equation}
From Eq.~\eqref{eq:linear_system_delta_phi_L_GP}, it follows that
\begin{equation} \label{eq:Castin_6_36}
	i \hbar \dot{b}_{k}(t)=\varepsilon_{k} b_{k}(t) \,,
\end{equation}
whose solution is
\begin{equation} \label{eq:coeff_b_time_evolution}
	b_{k}(t)=b_{k}(0)e^{-i \varepsilon_{k}t/\hbar} \,.
\end{equation}

In summary, when the spectrum is real, the expression for $\delta\phi$ in Eq.~\eqref{eq:delta_phi_real_espectra_first_row} corresponds to the conventional Bogoliubov expansion, is manifestly invariant under the conjugation transformation discussed in Sec.~\ref{sec:eigenbasis_expansion}, and allows for a convenient identification of the $u$ and $v^*$ components with positive and negative eigenvalues, respectively.

\subsection{Imaginary spectrum}
\label{sec:imaginary_spectrum}

When the spectrum acquires imaginary components, the standard mode expansion fails because the normalization condition in Eq.~\eqref{eq:normalization_condition_real_eigenvalue} cannot be satisfied, as the inner product vanishes even for $k' = k$; namely, $\braket{ u_{k} | u_{k}} - \braket{ v_{k} | v_{k}} = 0$~\footnote{From Eq.~(2.23) of Ref.~\cite{fetter1972}, which we rewrite as 
$(\varepsilon_k - \varepsilon_{k'}^*)
(\braket{u_k | u_{k'}} - \braket{v_k | v_{k'}})=0$, it can be inferred that, when the spectrum has an imaginary component, the second term must vanish identically. See also, e.g., Refs. \cite{wu2003,nakamura2008}.}.
The origin of this apparent failure is that one has to properly define the left eigenvectors for this case.
Considering that now $\varepsilon_{k}^{*} = -\varepsilon_{k}$, spectral property \ref{item:p1} implies that the right eigenvectors can be written as
\begin{equation}
\ket{\mathrm{R}_{k}} = \begin{pmatrix} \ket{u_{k}} \\ \ket{v_{k}} \end{pmatrix},
\, \begin{pmatrix} \ket{u_{k}^{*}} \\ \ket{v_{k}^{*}} \end{pmatrix},
\end{equation}
with eigenvalues $\pm\varepsilon_{k}$, respectively. The corresponding left eigenvectors can be chosen as (see spectral property~\ref{item:p4})
\begin{equation}
\ket{\mathrm{L}_{k}} = \mathcal{N}_{k}^{*} \begin{pmatrix} -\ket{v_{k}^{*}} \\ \ket{u_{k}^{*}} \end{pmatrix},\,\, 
\mathcal{N}_{k} \begin{pmatrix} -\ket{v_{k}} \\ \ket{u_{k}} \end{pmatrix},
\end{equation}
where the factor $\mathcal{N}_{k}$ is introduced to ensure that the normalization condition,
\begin{equation} \label{eq:normalization_condition_purely_imaginary_eigenvalue}
	\braket{\mathrm{L}_{k} | \mathrm{R}_{k^\prime}} = \mathcal{N}_{k}^{*} \left[ \braket{u_{k} | v_{k'}} - \braket{v_{k} | u_{k'}} \right]=\delta_{k,k'} \,,
\end{equation}
with $\mathcal{N}_j = -i$, is satisfied for both $\pm\varepsilon_{k}$.
Then, the expansion for $\delta\phi$ can be conveniently written as
\begin{equation} \label{eq:analog_of_eq:11_PR98_for_purely_imaginary_epsilon}
	\delta \phi(\bm{r}, t) = \sum_{k} c_{k}(t) u_{k}(\bm{r}) + d_k^{*}(t) u_k^{*}(\bm{r})\,.
\end{equation}
By setting $\mathrm{Im}(\varepsilon_{k}) > 0$, the first term can be identified with positive imaginary eigenvalues, and the second one with negative imaginary values, in analogy with the real case discussed above.
The expansion coefficients are given by  
\begin{align}
c_{k}(t) &\equiv i\left(\braket{u_{k} | \delta\phi^{*}} - \braket{v_{k} | \delta\phi}\right),
\\
d_{k}(t) &\equiv -i\left(\braket{u_{k} | \delta\phi} - \braket{v_{k} | \delta\phi^{*}}\right)\,.
\end{align}
These modes belong to an imaginary spectrum and evolve in time according to [see Eqs.~\eqref{eq:Castin_6_36} and~\eqref{eq:coeff_b_time_evolution} for a similar procedure]
\begin{align} 
\label{eq:coeff_c_time_evolution}
c_{k}(t) &= c_{k}(0) e^{\mathrm{Im}(\varepsilon_{k})t/\hbar},
\\
\label{eq:coeff_d_time_evolution}
d_{k}(t) &= d_{k}(0) e^{-\mathrm{Im}(\varepsilon_{k})t/\hbar},
\end{align}
indicating that the coefficients $c_{k}(t)$ grow exponentially while the coefficients $d_{k}(t)$ decay exponentially---a signature of dynamical instability.

Note that, with the above prescriptions, the expression for $\delta\phi$ in Eq.~\eqref{eq:analog_of_eq:11_PR98_for_purely_imaginary_epsilon} differs from the conventional Bogoliubov expansion \eqref{eq:delta_phi_real_espectra_first_row} and is not manifestly invariant under the conjugation transformation. However, we remark that this is just a matter of convention; the advantage of this form is that it explicitly separates positive and negative imaginary components.

\section{Quasiparticle projection method}
\label{sec:qpm}

The above expansion applies only to the linear regime of small-amplitude fluctuations---much smaller than the macroscopic occupation of the initial stationary solution---where they are treated as non-interacting among themselves. The solutions in Eqs.~\eqref{eq:coeff_b_time_evolution}, \eqref{eq:coeff_c_time_evolution}, and \eqref{eq:coeff_d_time_evolution} can describe their behavior at short times, but eventually break down when nonlinear effects come into play. Nevertheless, considering that the biorthogonal basis provided by the right and left eigenvectors of $\mathcal{L}$ constitutes a complete basis, one can generalize the usual Bogoliubov expansion to obtain a proper expansion valid at any time. This approach, known as the quasiparticle projection method, was originally proposed by Morgan \textit{et al.} in Ref.~\cite{morgan1998} for a real spectrum (see also Ref.~\cite{martinez-garaot2018}). 

Here, we generalize the analysis to the case in which $\mathcal{L}$ possesses a generic complex spectrum. This allows for a proper expansion around dynamically-unstable stationary solutions of the GP equation. 

\subsection{General case}
\label{sec:general_case}

Instead of Eq. \eqref{eq:small_deviation_from_steady}, we start by writing
\begin{equation} 
	\phi(\bm{r},t) = e^{-i\mu t/\hbar} \left[ b_{0}(t)\, \phi_{0}(\bm{r}) + \delta{\phi}(\bm{r},t) \right]  \,,
    \label{eq:phi_t}
\end{equation} 
where $b_{0}(t)$ accounts for a time-dependent population of the initial stationary state, in contrast to the usual Bogoliubov approach, where it is considered constant
\footnote{For uniformity of notation, we denote the coefficient of $\phi_0$ as $b_0(t)$, in contrast to the definition $1 + b_0(t)$ used in Refs. \cite{morgan1998,martinez-garaot2018}.}.
The above expression~\eqref{eq:phi_t} corresponds to the general solution of the time-dependent GP equation~\eqref{eq:GPE} with initial conditions $\phi(\bm{r},0) = b_{0}(0)\phi_{0}(\bm{r}) + \delta{\phi}(\bm{r},0)$. 
The specific expression for $b_{0}(t)$ will be presented below.
The contribution of the fluctuations can be expanded as in Eq.~\eqref{eq:expansion_left_right_eigenvec},
\begin{equation} \label{eq:expansion_delta_phi}
	\ket{\delta\Phi} = \sum_{j} b_{j}(t) \ket{\mathrm{R}_{j}}, 
\end{equation}
namely [see Eq.~\eqref{eq:psiR}],
\begin{align}
\begin{pmatrix}
		\delta\phi(\bm{r},t)\\
		\delta\phi^{*}(\bm{r},t)
	\end{pmatrix}
    =& \sum_{j}b_{j}(t)\begin{pmatrix} u_{j}(\bm{r}) \\ v_{j}(\bm{r}) \end{pmatrix},
\end{align}
where the expression for the coefficients $b_{j}$ can be obtained self-consistently using the completeness relation~\eqref{eq:completeness}.

Note that, although the eigenvectors $\left(\ket{u_{j}},\, \ket{v_{j}}\right)^\top$ obtained from the direct (possibly numerical) diagonalization of $\mathcal{L}$ are orthogonal to the zero-energy mode (see Appendix~\ref{app:zeroenergy}), their components $\ket{u_j}$ and $\ket{v_j}$ are not necessarily orthogonal to $\ket{\phi_{0}}$ and $\ket{\phi_{0}^{*}}$ individually. 
To ensure this, one can apply the substitutions $\ket{u_j}\to \ket{u_j}-\ket{\phi_{0}}\braket{\phi_{0} | u_j}$, and $\ket{v_j}\to \ket{v_j}-\ket{\phi_{0}^*}\braket{\phi_{0}^* | v_j}$ \cite{morgan1998}, with a similar transformation for the left eigenvectors. This is useful for encoding the entire condensate component in the expansion coefficient $b_{0}(t)$ and thus avoid double-counting
the condensate contribution to the mode expansion.
Under these assumptions, we then have
\begin{align}
\label{eq:b_0}
	b_{0}(t) &= \braket{\phi_{0} | \phi(t)}e^{i\mu t/\hbar},
    \\
    b_{j}(t) &= \braket{u_{j}^\textrm{L} | \phi(t)}e^{i\mu t/\hbar} + \braket{v_{j}^\textrm{L} | \phi^{*}(t)}e^{-i\mu t/\hbar}.
    \label{eq:b_j}
\end{align}
Eventually, the evolved wave function \eqref{eq:phi_t} can be expanded as
\begin{equation}
    \phi(\bm{r},t) = e^{-i\mu t/\hbar} \Big[ b_{0}(t)\, \phi_{0}(\bm{r}) + \sum_{j}b_{j}(t) u_{j}(\bm{r}) \Big],
    \label{eq:phi_t_reconst}
\end{equation}
where we recall that the index $j$ runs over all eigenvalues.
In general, it is also convenient to introduce distinct notations for the coefficients associated with the real and complex sectors of the spectrum. In particular, in the remainder of this paper, we will use the symbol $b_j$ for the coefficients associated with real eigenvalues, and $c_j$ and $d_j$ for those associated with complex eigenvalues with positive and negative imaginary components, respectively. See the discussion below.

\subsection{Spectrum of the form $\varepsilon = \pm \sqrt{f}$}
\label{sec:real_imaginary_spectrum}

Here, we consider the case in which the spectrum of $\mathcal{L}$ can be divided into real and imaginary sectors. This situation naturally arises when the condensate spectrum takes the form $\varepsilon = \pm \sqrt{f}$, where $f$ is a real-valued function that can be either positive or negative, resulting in real or imaginary energy eigenvalues, respectively. Such a spectrum is found, for example, in a uniform condensate at rest, either in the simple case of a negative scattering length \cite{castin2001} or in the presence of dipolar interactions \cite{blakie2020,alana2024}.

In this case, the contribution of fluctuations can be expanded as (see Sec.~\ref{sec:realL})
\begin{align}
	\delta\phi(\bm{r},t) =& \sum_{\varepsilon_{j}\in\mathbb{R}}b_{j}(t)u_{j}(\bm{r})+b_{j}^{*}(t)v_{j}^{*}(\bm{r})
	\nonumber\\
	& + \sum_{\varepsilon_{j}\in i\mathbb{R}} c_{j}(t) u_{j}(\bm{r}) + d_{j}^{*}(t) u_{j}^{*}(\bm{r}),
    \label{eq:reconstructed_box}
\end{align}
which accounts for the contributions from the real and imaginary sectors of the spectrum. 
Proceeding as in Sec.~\ref{sec:general_case}, the expression for the coefficients $b_{j}$, $c_{j}$, and $d_{j}$ can be obtained self-consistently from the completeness relation \eqref{eq:completeness}, yielding
\begin{align} 
	b_{j}(t) & = \braket{u_{j} | \phi(t)}e^{i\mu t/\hbar} - \braket{v_{j} | \phi^{*}(t)}e^{-i\mu t/\hbar} \,,
	\label{eq:coeff_b}
    \\
	c_{j}(t) & = \mathcal{N}_{j}^{*}\!\left( \braket{u_{j} | \phi^{*}(t)}e^{-i\mu t/\hbar} - \braket{v_{j} | \phi(t)}e^{i\mu t/\hbar}\right),
	\label{eq:coeff_c}
    \\
	d_{j}(t)  & = \mathcal{N}_{j}\left( \braket{u_{j} | \phi(t)}e^{i\mu t/\hbar} - \braket{v_{j} | \phi^{*}(t)}e^{-i\mu t/\hbar}\right).
    \label{eq:coeff_d}
\end{align}

Now that the formalism has been established, we proceed to illustrate its use in practical examples by analyzing the evolution of the condensate wave function in the presence of a dynamical instability. We remark that the method is not predictive, as it relies on a known solution of the GP equation provided as input.
Nonetheless, it serves as a powerful tool for analyzing the overall behavior and emergence of dynamical instabilities at arbitrary times. 
We recall that linear-stability analysis determines whether the system is dynamically unstable and which modes grow exponentially---a characterization intrinsically limited to the initial, linear regime. However, the present approach allows one to track how these modes contribute to the dynamics at any given time, to identify which nominally unstable modes dominate the evolution, and to establish a direct link between the short-time linear regime and the long-time formation of macroscopic density structures. Therefore, this framework provides a consistent description of the system even after it has left the linear regime, where linear stability analysis no longer applies.

In a broad sense, the approach can be seen as analogous to the Fourier analysis of a given signal---in this case, the condensate wave function---carried out using a complete basis of normal modes of the system under consideration.

\section{Proof of concept: Condensate~in~a~box}
\label{sec:proof_of_concept}

As a proof of principle for how the generalized quasiparticle projection method works in practice, we consider the case of a one-dimensional condensate confined in a box of size $L$ with periodic boundary conditions (PBCs). 
It is convenient to work in dimensionless units by introducing an arbitrary length scale $\ell$ and the corresponding energy scale $\hbar\omega_{\ell}$, with $\ell = \sqrt{\hbar/(m\omega_{\ell})}$, thereby effectively setting $\hbar = 1 = m$ in the dimensional GP equation.
Here, we consider stationary solutions of the form
\begin{equation}
\phi_0 = \sqrt{N/L}\,e^{iqx},
\label{eq:factorization}
\end{equation}
corresponding to a uniform condensate moving with momentum $q$.
These solutions are characterized by a chemical potential $\mu = q^2/2 + g\rho_0$, where the density $\rho_0 \equiv |\phi_0|^2 = N/L$ represents the number of atoms per unit length. Therefore, it is convenient to employ the factorization discussed in Appendix~\ref{sec:Bloch_factorization} [see Eq.~\eqref{eq:small_deviation_from_steady_3}]. Note that, apart from the substitution $\mu t/\hbar \to \mu t/\hbar - qx$, all the discussion and results of the previous section hold.
The Bogoliubov functions $u$ and $v$ also take the form of plane waves, labeled by the corresponding wave vector $k$ \cite{castin2001},
\begin{equation}
	\begin{pmatrix} \label{eq:eigenvectors_condensate_box_eikr_over_L}
		u_{k}(x)\\v_{k}(x)
	\end{pmatrix}
	= \frac{e^{ikx}}{\sqrt{L}}
	\begin{pmatrix}
		U_k\\V_k
	\end{pmatrix},
\end{equation}
and they are eigenstates of the following Bogoliubov matrix,
\begin{equation} \label{eq:L_GP_box}
	\mathcal{L}=
	\begin{pmatrix}
		k^{2}/2 + kq + g\rho_{0} & g\rho_{0}\\
		-g\rho_{0} & -\left[k^{2}/2 - kq + g\rho_{0}\right]
	\end{pmatrix},
\end{equation}
whose eigenvalues,
\begin{equation} \label{eq:Bogoliubov_spectrum}
	\varepsilon_k^\pm = kq \pm \sqrt{ \frac{k^2}{2} \left( \frac{k^2}{2} + 2g\rho_0 \right) },
\end{equation}
constitute the well-known Bogoliubov spectrum. Owing to PBCs, the allowed values of the momentum are
\begin{equation} 
\label{eq:k_equal_n_pi_over_2L} 
k \equiv k_n = {2\pi}n/{L}, 
\end{equation} 
with $n=0,\pm1,\pm2\,\ldots$. 
Interestingly, for attractive interactions ($g < 0$), the spectrum acquires an imaginary component for $k < k^{*} \equiv 2\sqrt{|g|\rho_0}$, signaling that the system becomes dynamically unstable. 
In the following, we discuss how the Bogoliubov quasiparticle approach applies in the cases of a condensate at rest ($q=0$) and a moving condensate ($q \neq 0$).

\subsection{Attractive condensate at rest}
\label{sec:bec_at_rest}

\begin{figure}[t!]
\centerline{\includegraphics[width=\columnwidth]{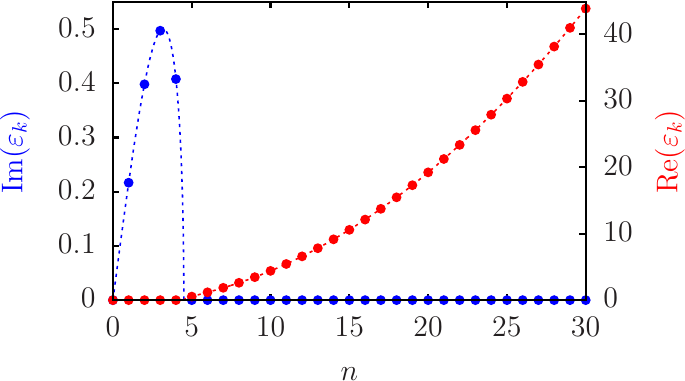}}
	\caption{Bogoliubov dispersion relation $\varepsilon_k^+$ [see Eq.~\eqref{eq:Bogoliubov_spectrum}] for a uniform BEC with $g<0$, shown here for positive values of the momentum wave vector $k$. For $k < k^{*}$, the spectrum is purely imaginary, while it becomes purely real for $k \ge k^{*}$. 
	Here, $gN=-10$ and $L=20$. 
    We recall that the wave vector $k$ has discrete values $k_n$, as defined in Eq.~\eqref{eq:k_equal_n_pi_over_2L}.
    }
	\label{fig:spectrum_box}
\end{figure}

\paragraph*{Bogoliubov spectrum.}
For $g < 0$ and $q = 0$, the spectrum,
\begin{equation} \label{eq:Bogoliubov_spectrum_q=0}
    \varepsilon_k^\pm = \pm \sqrt{ \epsilon_k \left( \epsilon_k + 2\beta \right) },
\end{equation}
is either real or imaginary, depending on whether $k \ge k^{*}$ or not.
We define $\beta \equiv g\rho_0$, and $\epsilon_k = k^2 / 2$ corresponds to the free-particle dispersion relation.
Here, $\varepsilon_k^+$ denotes the principal value of the square root, which is defined with a non-negative real (positive imaginary) part when the argument is positive (negative).
As an example, we set $L = 20$ and $gN = -10$, yielding $k^{*} = \sqrt{2}$; for the mode index $n$, this condition reads $n^{*} = 5$. Figure~\ref{fig:spectrum_box} shows the corresponding spectrum.

We now apply the formalism developed in Sec.~\ref{sec:real_imaginary_spectrum}. In particular, by properly normalizing the eigenvectors as described therein [see Eqs. \eqref{eq:normalization_condition_real_eigenvalue} and \eqref{eq:normalization_condition_purely_imaginary_eigenvalue}], we obtain explicit expressions for $U_k$ and $V_k$ \footnote{The orthogonality for $k^{\prime} \neq k$ is guaranteed by the plane wave component $\exp(ikx)$, following from $\int_{-L/2}^{L/2} \exp\{i({k} - {k^{\prime}})x\} \, dx = L\, \delta_{{k}{k^{\prime}}}$.}. 
To simplify the notation, we introduce $\zeta \equiv \beta + \epsilon_k + \varepsilon_k^+$.
Depending on whether the excitation energy $\varepsilon_k$ is real or imaginary, the expressions for $U_k$ and $V_k$ take different forms:
for $\varepsilon_k\in \mathbb{R}$,
\begin{equation}
	U_k = -\frac{\zeta}{\beta \sqrt{ {\zeta^2}/{\beta^2}-1}}, 
    \quad 	
    V_k = \frac{1}{\sqrt{{\zeta^2}/{\beta^2}-1 }}  \,,
\end{equation}
and, for $\varepsilon_k\in i\mathbb{R}$,
\begin{equation}
    U_k = -\frac{\zeta}{\sqrt{2i\beta\varepsilon_k^+}}, 
    \quad 	
    V_k = \sqrt{\frac{\beta}{2i\varepsilon_k^+}}.
\end{equation}

\begin{figure}
\centerline{\includegraphics[width=\columnwidth]{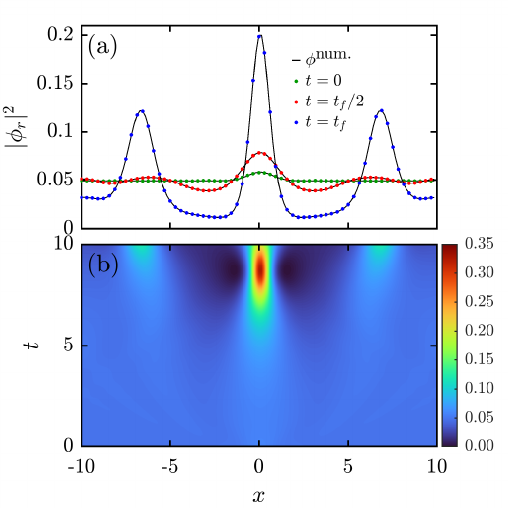}}
	\caption{Evolution of a uniform condensate at rest with $g<0$. (a) Comparison between the numerically evolved density (solid line) and the reconstructed one (dots) at selected times $t = 0$, $t_f/2$, and $t_f$. (b) Density plot of the evolved density as a function of time. See text for additional details.
    }
	\label{fig:reconstructed_box}
\end{figure}

\paragraph*{GP evolution and quasiparticle projection analysis.}
Initially, the condensate is prepared in the state $\phi(x,t=0) = \phi_0 + \delta\phi(x)$, with the perturbation $\delta\phi(x)$ in the form of a Gaussian centered at $x = 0$ [see black dots in Fig.~\ref{fig:reconstructed_box}(a)]. 
In one dimension, a condensate can support solitonic solutions, so it is not expected to collapse for $g < 0$~\footnote{See, e.g., Ref.~\cite{dalfovo1999} and Chapter 7 of Ref.~\cite{pethick2008} (and references therein).}. 
The choice of this initial condition effectively imprints a seed that triggers the formation of localized---yet nonstationary---structures. 
The form of $\delta\phi(x)$ also affects the initial population of Bogoliubov modes, whose amplitude decays with $k$ for the present choice.
We then evolve the condensate wave function according to the GP equation~\eqref{eq:GPE} until a localized structure forms. At this point, the dynamical instability has driven the system out of the linear regime, and nonlinear effects dominate.
In the present case, we consider evolutions up to a final time $t_f = 10$, expressed in the dimensionless units defined above.

The evolution of the density distribution is shown in Fig.~\ref{fig:reconstructed_box} at selected times in panel (a) and as a density plot over time in panel (b). 
In Fig.~\ref{fig:reconstructed_box}(a), we compare the density profile with the expression for the \emph{reconstructed wave function} obtained from Eqs.~\eqref{eq:small_deviation_from_steady} and \eqref{eq:reconstructed_box}, indicated here as $\phi_r(x,t)$. Excellent agreement is observed at different times.
Note that the expansion of $\delta\phi(x)$ in Eq.~\eqref{eq:reconstructed_box} is truncated by including a limited number of Bogoliubov modes, up to $k_n=k_{\tilde{n}}$, which depends on time. 
In particular, here we use $\tilde{n}=8$ for $t=0$ and $t_f/2$, and $\tilde{n}=16$ for $t=t_f$. The corresponding fidelity $\mathcal{F}\equiv|\braket{\phi_r(t)|\phi(t)}|$  deviates from unity by less than $0.001\%$.
This demonstrates that the Bogoliubov quasiparticle projection method can provide a very accurate reconstruction of the evolved wave function deep in the nonlinear regime---even with a restricted set of modes.

\begin{figure}
\centerline{\includegraphics[width=0.9\columnwidth]{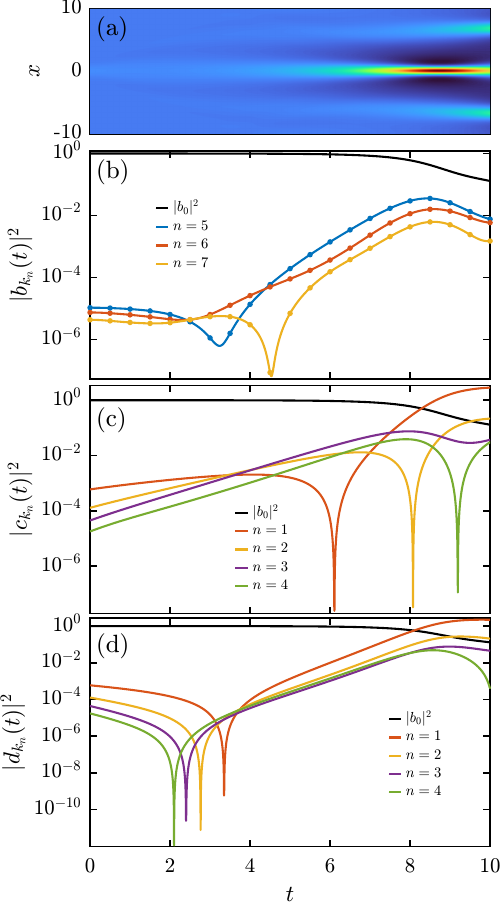}}
    \caption{
    Evolution of the Bogoliubov coefficients for a uniform condensate at rest with $g < 0$. (a) Density plot over time, as in Fig.~\ref{fig:reconstructed_box}(b), shown here for reference. 
    (b--d) Time evolution of the modulus squared of the coefficients $b_k(t)$, $c_k(t)$, and $d_k(t)$, respectively. Here $k$ takes discrete values $k = k_n$, as defined in Eq.~\eqref{eq:k_equal_n_pi_over_2L}.
    The coefficients $b_k$ are associated with the real sector of the spectrum ($n = 5, 6, 7$); the solid line (dots) corresponds to the $+$ ($-$) sign in Eq.~\eqref{eq:Bogoliubov_spectrum_q=0}.
    The coefficients $c_k$ and $d_k$, associated with the imaginary sector ($n = 1, 2, 3, 4$), correspond to positive and negative imaginary parts, respectively.
    In panels (b--d), we also show the condensate amplitude $b_0$ for reference.
    See text for additional details.
    }
	\label{fig:coefs_evolution_box}
\end{figure}

In Fig.~\ref{fig:coefs_evolution_box}, we show the evolution of some of the coefficients in expression~\eqref{eq:reconstructed_box}, obtained from the definitions in Eqs.~\eqref{eq:b_0} and \eqref{eq:coeff_b}--\eqref{eq:coeff_d}.
We recall that the coefficients $b_k$ [in panel (b)] are those associated with real eigenvalues, and $c_k$ and $d_k$ [panels (c) and (d)] correspond to complex eigenvalues with positive and negative imaginary components, respectively.
In the top panel, we include the same density plot shown in Fig.~\ref{fig:reconstructed_box}(b) (with the axes interchanged) for reference.

In the linear regime, we therefore expect the coefficients to roughly follow the behavior dictated by Eqs.~\eqref{eq:coeff_b_time_evolution}, \eqref{eq:coeff_c_time_evolution}, and \eqref{eq:coeff_d_time_evolution}. Initially, the coefficients $b_k$ have an almost constant square modulus, with a small decay at later times due to the transfer of population between different modes; this arises from norm conservation and is not accounted for in the usual linear approximation. The coefficients $c_k$, on the other hand, exhibit an exponentially growing square modulus, whereas $d_k$ are exponentially suppressed.

\begin{figure}
    \centerline{\includegraphics[width=0.9\columnwidth]{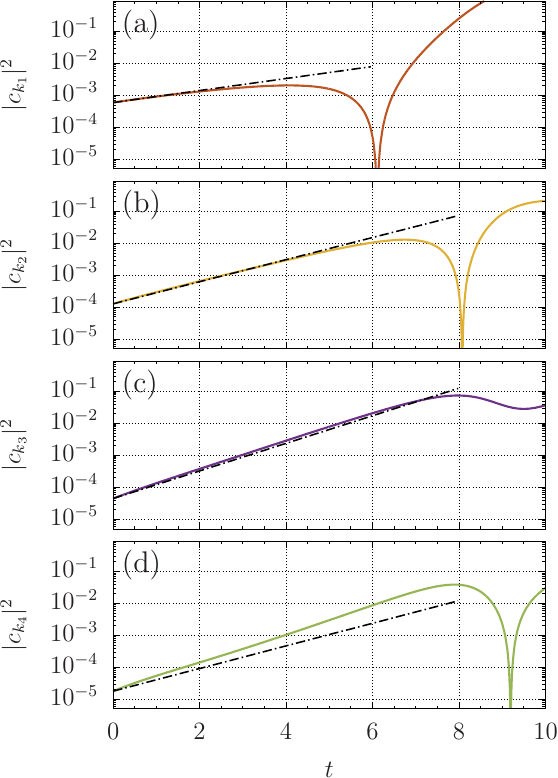}} 
	\caption{Evolution of the population of the same dynamically unstable modes as in Fig.~\ref{fig:coefs_evolution_box}(c). Each mode is displayed in a separate panel and compared with the prediction of the linear stability analysis, $|c_{k}(t)|^2 = |c_{k}(0)|^2 \exp{2\,\mathrm{Im}(\varepsilon_{k})t/\hbar}$, shown as dashed-dotted lines. The results are shown for $k=k_n$, as defined in Eq.~\eqref{eq:k_equal_n_pi_over_2L}, with $n=1,2,3,4$.
    }
	\label{fig:c_t}
\end{figure}

\paragraph*{Linear regime and nonlinear mixing.}
The behavior of the coefficients $|c_k(t)|^2$ introduced in Fig.~\ref{fig:coefs_evolution_box}(c) is also shown in Fig.~\ref{fig:c_t} (one coefficient per panel), where we compare them with the prediction of the linear stability analysis, $|c_k(t)|^2 = |c_k(0)|^2 \exp\{2\,\mathrm{Im}(\varepsilon_k)t/\hbar\}$~[see Eq.~\eqref{eq:coeff_c_time_evolution}], shown as dashed-dotted lines.

With the present choice of initial conditions---the form of the perturbation $\delta\phi(x,t=0)$---the initial population of the modes decreases with increasing $n$. 
From the figure, we see that the first three modes start growing exponentially, at a rate set by the imaginary component of the corresponding eigenvalue, in agreement with the analytical prediction.  

Only the lowest populated mode exhibits a noticeable deviation from the expected growth rate. We attribute this to the fact that this mode, being the least populated, can experience nonlinear effects due to the presence of the other modes---an effect not accounted for by the linear theory. Indeed, we have verified that, when only individual Bogoliubov modes are initially populated, their exponential growth matches the linear stability prediction.

The nonlinear coupling between Bogoliubov modes causes the first mode, which has the lowest growth rate, to leave the linear (exponentially growing) regime at around $t = 2$---earlier than the other three modes, which have higher growth rates and reach this point between $t=6$ and $t=8$. 
Even at earlier times, however, the nonlinear coupling between the modes becomes dominant: several modes already deviate from the linear regime prediction. For example, some of the coefficients $b_k$ and $d_k$ also develop an exponentially growing behavior, as shown in Figs.~\ref{fig:coefs_evolution_box}(b) and (d). 
At later times, the nonlinear mixing between modes ultimately gives rise to localized macroscopic structures, which are a direct manifestation of the dynamical instability.

Altogether, the present method provides a complete and robust mode expansion that remains meaningful beyond the linear regime.
The evolution of the coefficients then provides a clear modal picture of nonlinear population transfer, revealing how mode mixing can amplify initially weak modes that only become relevant at later times, thereby characterizing how the instability develops at the macroscopic level.

\subsection{Attractive condensate with nonzero momentum}
\label{sec:moving_bec}

\begin{figure}[t!]
    \centerline{\includegraphics[width=\columnwidth]{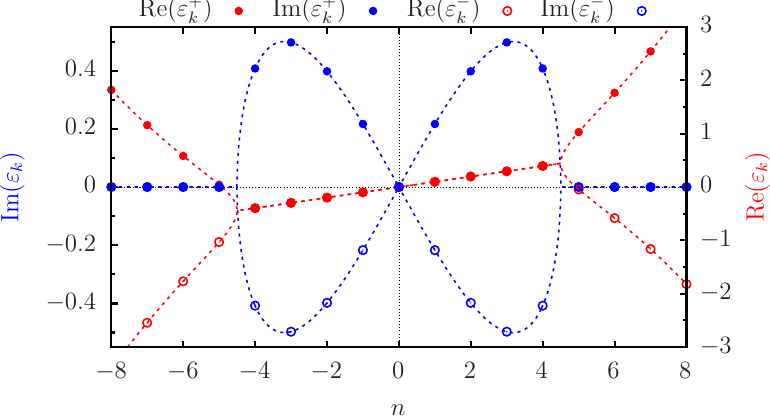}}
	\caption{Bogoliubov dispersion relation $\varepsilon_k^\pm$ [see Eq.~\eqref{eq:Bogoliubov_spectrum}] for a uniform BEC with $g<0$ and momentum $q = 2\pi/L$. The other parameters are the same as in Fig.~\ref{fig:spectrum_box}.
    }
	\label{fig:spectrum_box_vel}
\end{figure}

We now consider a moving condensate ($q \neq 0$). In this case, the spectrum becomes complex for $k < k^{*}$ rather than purely imaginary, and an explicit analytic solution for the Bogoliubov eigenvectors is no longer straightforward. 
It is therefore convenient to refer to the general expansion of Sec.~\ref{sec:general_case} and rely on a simple---though numerical---solution \footnote{Note that, in general, diagonalization routines of biorthogonal matrices return by default both right and left eigenvectors.}. 
We show the resulting spectrum in Fig.~\ref{fig:spectrum_box_vel}; note that the spectrum is no longer symmetric in $k$ due to the nonzero momentum.
\begin{figure}
	\centerline{\includegraphics[width=\columnwidth]{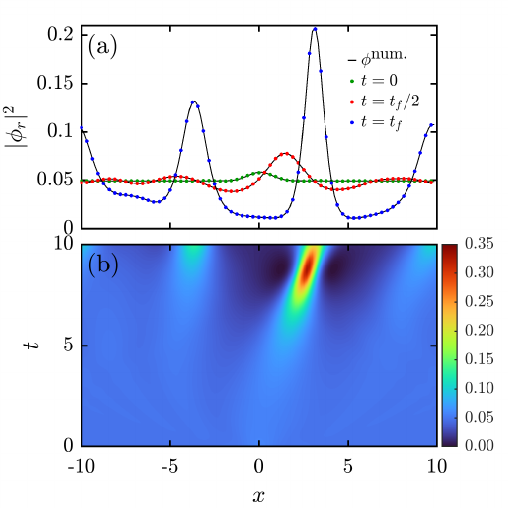}}
	\caption{Evolution of a uniform condensate with momentum $q = 2\pi/L$ and $g<0$. (a) Comparison between the numerically evolved density (solid line) and the reconstructed one (dots) at times $t = 0$, $t_f/2$, and $t_f$. (b) Density plot of the evolved density as a function of time. See text for additional details.}
	\label{fig:reconstructed_box_vel}
\end{figure}

This numerical, more general approach also allows testing the alternative expression in Eq.~\eqref{eq:expansion_delta_phi} for the Bogoliubov quasiparticle expansion. 
It is worth remarking that the expressions for the expansion coefficients $b_{0}(t)$ and $b_{j}(t)$ in Eqs.~\eqref{eq:b_0} and \eqref{eq:b_j} must be modified accordingly to account for the factorization we are using in this section [see Eqs.~\eqref{eq:factorization} and \eqref{eq:small_deviation_from_steady_3}].
This can be implemented in those equations by the replacement
\begin{equation}
\ket{\phi(t)} \to e^{-iqx}\ket{\phi(t)},
\end{equation}
which leads, for instance, to
\begin{equation}
\braket{u_{j}^\textrm{L} | \phi(t)} \to \int u_{j}^{\textrm{L}^*}\!\!(x)e^{-iqx}\phi(x,t)dx,
\end{equation}
and similarly for all other terms.

\begin{figure}
	\centerline{\includegraphics[width=0.9\columnwidth]{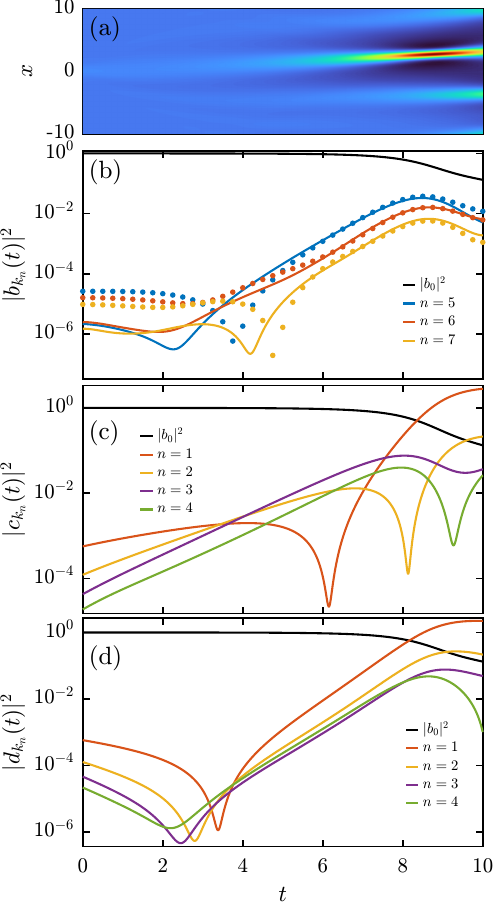}}
	\caption{Evolution of the Bogoliubov coefficients for a uniform, moving condensate, with $g < 0$. (a) Density plot over time, as in Fig.~\ref{fig:reconstructed_box_vel}(a), shown here for reference. (b--d) Time evolution of the modulus squared of the coefficients $b_k(t)$, $c_k(t)$, and $d_k(t)$, respectively. In panel (b), the solid line (dots) corresponds to the $+$ ($-$) sign in Eq.~\eqref{eq:Bogoliubov_spectrum_q=0}. Refer to the caption of Fig.~\ref{fig:coefs_evolution_box} for a more detailed description.
    }
	\label{fig:coefs_evolution_box_vel}
\end{figure}

In the following, as done in the previous case, we denote by $b_k$ the coefficients associated with real eigenvalues, and by $c_k$ and $d_k$ those that correspond to complex eigenvalues with positive and negative imaginary parts, respectively.
For illustration purposes, we consider the case $q = 2\pi/L$. By performing the same analysis as in the previous section, we obtain the results shown in Figs.~\ref{fig:reconstructed_box_vel} and \ref{fig:coefs_evolution_box_vel}.
We find that the reconstruction procedure remains accurate also in the moving case, and that the evolution of the coefficients closely resembles that of the $q=0$ configuration. In particular, we observe the same distinction between real- and complex-eigenvalue sectors, and the dominant unstable modes exhibit a similar exponential growth. This confirms the robustness of the method beyond the linear regime.

\section{Conclusions}
\label{sec:conclusions}

We have presented a general formalism for performing a time-dependent Bogoliubov analysis of dynamically unstable BECs, extending the quasiparticle projection method of Morgan \textit{et al.} \cite{morgan1998} to the case of a generic complex spectrum. By introducing the appropriate left eigenvectors associated with each regime, one can construct a biorthogonal basis of single-frequency modes that enables a complete mode decomposition. This framework provides a natural extension of the Bogoliubov theory beyond the stable regime, remaining valid even when the excitation spectrum includes complex or imaginary eigenvalues.

The method provides a powerful tool for analyzing the emergence and development of dynamical instabilities at arbitrary times based on a known solution of the GP equation.
In particular, it allows one to clarify the validity of the linearity hypothesis for a given set of initial conditions, to identify the modes that contribute most significantly to the instability, and to connect the long-time nonlinear dynamics with the short-time exponential growth of unstable modes predicted by linear stability analysis.
In a broader sense, this approach can be regarded as analogous to a Fourier analysis of a given signal---here represented by the condensate wave function---carried out using the complete set of Bogoliubov modes appropriate to the system. This makes it possible to reconstruct the condensate evolution under arbitrary perturbations, providing a consistent description of unstable dynamics in ultracold gases and, potentially, in other nonlinear systems governed by similar mean-field equations.

As a proof of concept, we have applied the method to a one-dimensional condensate with attractive interactions, dynamically unstable and evolving into nonstationary localized structures seeded by small perturbations.
The Bogoliubov quasiparticle projection method is shown to accurately reconstruct the evolved wave function deep in the nonlinear regime, even with a limited number of modes. This opens up interesting perspectives for applications to more complex scenarios, such as dynamical instabilities in optical lattices \cite{wu2001,wu2003,fallani2004,modugno2004,desarlo2005}, the roton instability leading to the formation of supersolid structures in dipolar condensates \cite{alana2024}, or vortex-array instabilities in superfluid counterflows \cite{giacomelli2023,hernandez-rajkov24}, to mention just a few examples.
From a formal perspective, the formulation of canonical equations for the mode amplitudes in the biorthogonal basis \cite{morgan1998} in the presence of a complex spectrum represents an additional interesting direction for future work.

\begin{acknowledgments}
We acknowledge support from the Basque Government through Grant No.~IT1470-22, from Grant No.~PID2021-126273NB-I00, funded by MCIN/AEI/10.13039/501100011033 and by “ERDF A way of making Europe”, and from the European Research Council through the Advanced Grant “Supersolids” (No. 101055319).
AI acknowledges financial support from the fellowship PIF22/136 of the UPV/EHU.
\end{acknowledgments}

\appendix
\section{Alternative factorizations}
\label{app:alt_factorization}

\subsection{Modulo-phase factorization}
\label{sec:phase_factorization}

This is a conventional factorization in which the wave function $\phi_{0}$ of the stationary state is decomposed in the product of its modulus (the square root of the stationary density $|\phi_{0}|=\sqrt{n_{0}}$) times a phase factor (see, e.g., Refs.~\cite{fetter1972,leonhardt2003}),
\begin{equation}
    \phi(\bm{r},t) = |\phi_{0}(\bm{r})|\exp{iS(\bm{r}) -i\mu t/\hbar},
\end{equation}
yielding
\begin{equation} 
\label{eq:small_deviation_from_steady_2}
\phi(\bm{r},t)	= e^{-i\mu t/\hbar} e^{iS(\bm{r})} \left[|\phi_{0}(\bm{r})| + \delta{\phi}(\bm{r},t) \right],
\end{equation}
and
\begin{equation} \label{eq:L_GP_2}
	\mathcal{L}=
	\begin{pmatrix}
		H_{\textrm{GP}}^{(S)}+g|\phi_0|^2 & g|\phi_0|^2\\
		-g|\phi_0|^2 & -\left[H_{\textrm{GP}}^{(S)}+g|\phi_0|^2\right]^*
	\end{pmatrix},
\end{equation}
with
\begin{equation} \label{eq:H_GP_2}
	H_{\textrm{GP}}^{(S)}\equiv -\frac{\hbar^2}{2m}(\nabla + i\nabla S(\bm{r}))^2  + V(\bm{r}) + g|\phi_0|^2-\mu.
\end{equation}

This decomposition has the advantage that the off-diagonal terms are real, but the momentum shift introduces a complex contribution in the kinetic term. Although the operator remains Hermitian, it is no longer real.

\subsection{Bloch factorization}
\label{sec:Bloch_factorization}

This factorization is convenient when the condensate is subject to a periodic potential or moving with nonzero momentum~(see, e.g., Refs.~\cite{wu2001,wu2003,modugno2004}). For simplicity, we consider here a one-dimensional system. In this case, stationary states take the form of Bloch waves, $\phi_{q}(x) = \varphi_{q}(x) e^{iqx}$, so that one can conveniently write
\begin{equation} 
\label{eq:small_deviation_from_steady_3}
\phi(x,t)	= e^{-i\mu t/\hbar} e^{iqx} \left[\varphi_{q}(x) + \delta{\varphi}(x,t) \right],
\end{equation}
so that
\begin{equation} \label{eq:L_GP_3}
	\mathcal{L}=
	\begin{pmatrix}
		H_{\textrm{GP}}^{(q)}+g|\varphi_{q}|^2 & g\varphi_{q}^2\\
		-g\varphi_{q}^{*2}& -\left[H_{\textrm{GP}}^{(q)}+g|\varphi_{q}|^2\right]^*
	\end{pmatrix},
\end{equation}
with
\begin{equation} \label{eq:H_GP_3}
	H_{\textrm{GP}}^{(q)}\equiv -\frac{\hbar^2}{2m}(\partial_{x} + iq)^2  + V(\bm{r}) + g|\varphi_{q}|^2-\mu.
\end{equation}

This decomposition has the advantage of being expressed in terms of the natural quantum number for a periodic system: the quasimomentum. Note that, unlike the previous case, the periodic components $\varphi_{q}(x)$ are generally characterized by an additional local phase factor (besides the factor $e^{iqx}$), so that the off-diagonal terms are, in general, not real.

\section{Orthogonalization to the zero-energy mode}
\label{app:zeroenergy}

It is straightforward to verify that the operator ${\cal L}$ admits the following right and left zero-energy modes \cite{morgan1998}:
\begin{equation}
     \ket{R_0} = \begin{pmatrix} \ket{\phi_{0}} \\ -\ket{\phi_{0}^{*}} \end{pmatrix}, \quad
     \ket{L_0} = \begin{pmatrix} \ket{\phi_{0}} \\ \ket{\phi_{0}^{*}}
     \end{pmatrix}.
\end{equation}
In general, the eigenvectors $\ket{R_{j}}=\left(\ket{u_{j}},\, \ket{v_{j}}\right)^\top$ and $\ket{L_{j}}=\left(\ket{u_{j}^L},\, \ket{v_{j}^L}\right)^\top$, associated with a nonzero eigenvalue $\epsilon_j \neq 0$, are orthogonal to the zero-energy modes defined above, in the sense of Eq.~\eqref{eq:biorthogonality2}:
\begin{align}
    \braket{\phi_{0} | u_{j}} + \braket{\phi_{0}^* | v_{j}} &= 0\,,
    \\
    \braket{u_{j}^L | \phi_{0}} - \braket{v_{j}^L | \phi_{0}^*} &= 0\,.
\end{align}

However, the components $u_j$ and $v_j$ are not necessarily orthogonal to $\phi_{0}$ and $\phi_{0}^{*}$ individually. 
To impose this additional condition---and avoid double counting of the condensate contribution in the mode expansion---one may apply the transformations~\cite{morgan1998}
\begin{align}
\ket{\tilde{u}_j} &= \ket{u_j}-\ket{\phi_{0}}\braket{\phi_{0} | u_j}, 
\\
\ket{\tilde{v}_j} &= \ket{v_j}-\ket{\phi_{0}^{*}}\braket{\phi_{0}^{*}| v_j},
\end{align}
with analogous transformations applied to the left eigenvectors.
It is straightforward to verify that this procedure leaves the biorthogonality condition invariant, 
and therefore holds for the transformed vectors,
\begin{equation}
    \braket{\tilde{u}_{j}^L | \tilde{u}_{j'}} + \braket{\tilde{v}_{j}^L | \tilde{v}_{j'}} = \delta_{j,j'}.
\end{equation}

\bibliography{biblio}

\end{document}